\documentclass{article}

\usepackage{PRIMEarxiv}

\usepackage[utf8]{inputenc} % allow utf-8 input
\usepackage[T1]{fontenc}    % use 8-bit T1 fonts
\usepackage{hyperref}       % hyperlinks
\usepackage{url}            % simple URL typesetting
\usepackage{booktabs}       % professional-quality tables
\usepackage{amsfonts}       % blackboard math symbols
\usepackage{amsmath}
\usepackage{nicefrac}       % compact symbols for 1/2, etc.
\usepackage{microtype}      % microtypography
\usepackage{lipsum}
\usepackage{graphicx}
\usepackage{subfigure}
\usepackage{mathpazo}
\usepackage{sectsty}
\usepackage{tikz}
\usepackage{algorithm}
\usepackage{algorithmicx}
\usepackage{algpseudocode}

\usepackage{booktabs}
\usepackage{subcaption}
\usepackage{multirow}
\usepackage{array}
\usepackage{color}
\newcommand{\change}[1]{\textcolor{black}{#1}}

\usepackage{pifont}% http://ctan.org/pkg/pifont

\graphicspath{{media/}}     % organize your images and other figures under media/ folder

\title{A nonlinear extension of parametric model embedding for dimensionality reduction in parametric shape design
}

\author{
  A. Serani$^{1,\star}$, G. Palma$^1$, and M. Diez$^1$\\
  $^1$National Research Council-Institute of Marine Engineering, Rome, Italy\\
  $^\star$\texttt{andrea.serani@cnr.it} \\
}

\begin{document}
% \begin{tikzpicture}[remember picture,overlay]
%    \node [rectangle, fill=cyan, fill opacity=0.5, anchor=north, minimum width=\paperwidth, minimum height=3cm, text width=\textwidth, align=center, text height=5ex, text depth=10ex, align=left] at (current page.north) {\sffamily\small 
%    \textbf{This is a preprint of the following article:}\\
%    A. Serani, T. P. Scholcz, V. Vanzi, A Scoping Review on Simulation-based Design Optimization in Marine Engineering: Trends, Best Practices, and Gaps. \textit{Archives of Computational Methods in Engineering}, 2024.\\
%    \textbf{The published article is available by following the DOI: \texttt{10.1007/s11831-024-10127-1}, which may differ from this preprint.}
%    };
% \end{tikzpicture}

\begin{tikzpicture}[remember picture,overlay]
   % Nodo per il riempimento con trasparenza
   \node [rectangle, fill=cyan, fill opacity=0.5, anchor=north, minimum width=\paperwidth, minimum height=3cm] at (current page.north) {};

   % Nodo separato per il testo, senza trasparenza
   \node [anchor=north, minimum width=\paperwidth, minimum height=3cm, text width=\textwidth, align=center, text height=5ex, text depth=15ex, align=left] at (current page.north) {
     \sffamily\small
     \textbf{This is a preprint submitted to:} \textit{International Journal for Numerical Methods in Engineering}
     % \textbf{This is a preprint of the following article:}\\
     % A. Serani and M. Diez, A Scoping Review on Simulation-based Design Optimization in Marine Engineering: Trends, Best Practices, and Gaps. \textit{Archives of Computational Methods in Engineering}, 2024.\\
     % \textbf{The published article is available by following the DOI: \texttt{10.1007/s11831-024-10127-1}, which may differ from this preprint.}
   };
\end{tikzpicture}

\maketitle

\begin{abstract}
Dimensionality reduction is essential in simulation-based shape design, where high-dimensional parameterizations hinder optimization, surrogate modeling, and systematic design-space exploration. Parametric Model Embedding (PME) addresses this issue by constructing reduced variables from geometric information while preserving an explicit backmapping to the original design parameters. However, PME is intrinsically linear and may become inefficient when the sampled design space is governed by nonlinear geometric variability. This paper introduces a nonlinear extension of PME, denoted NLPME. The proposed framework preserves the defining principle of PME---geometry-driven latent variables and parameter-mediated reconstruction---while replacing the linear reduced subspace with a nonlinear latent representation. Geometry is not reconstructed directly from the latent variables; instead, the latent representation is decoded into admissible design parameters, and the corresponding geometry is recovered through a forward parametric map. The method is assessed on a bio-inspired autonomous underwater glider with a 32-dimensional parametric shape description and a CAD-based geometry-generation process. NLPME reaches a 5\% reconstruction-error threshold with \(N=5\) latent variables, compared with \(N=8\) for linear PME, and a 1\% threshold with \(N=9\), compared with \(N=15\) for PME. Comparison with a deep autoencoder shows that most of the nonlinear compression gain can be retained while preserving an explicit backmapping to the original design variables. The results establish NLPME as a compact, admissible, and engineering-compatible nonlinear reduced representation for parametric shape design spaces.
\end{abstract}

% keywords can be removed

\section{Introduction}

High-dimensional parametric descriptions are pervasive in contemporary engineering design. In simulation-based shape design, increasingly flexible geometric parameterizations, high-fidelity numerical solvers, and data-driven workflows have made it possible to explore richer and more realistic design spaces than was previously feasible. At the same time, this progress has intensified one of the central computational bottlenecks of engineering design, namely the \emph{curse of dimensionality} \cite{bellman1957-CS}: as the number of design variables grows, optimization becomes more fragile, surrogate models require increasingly large training sets, and uncertainty quantification rapidly loses tractability and robustness. Dimensionality reduction is therefore not merely a convenience, but a representation-level strategy for making high-dimensional design spaces computationally manageable \cite{serani2024survey}.

This issue is particularly acute in parametric shape optimization, where admissible geometries are generated through finite-dimensional design vectors, but the resulting family of shapes may still exhibit nontrivial structure in the geometric space \cite{manzoni2022shape}. Classical dimensionality-reduction approaches in this setting are predominantly linear. Proper orthogonal decomposition \cite{toal2010geometric,zhang2018multidisciplinary,liu2024aerodynamic}, Karhunen--Lo\`eve expansions \cite{diez2015-CMAME,chang2023research}, and their discrete implementations through principal component analysis \cite{yonekura2014shape,dagostino2020-OE,harries2021application} have been widely used to compress shape spaces and define reduced design variables for simulation-based optimization. Their appeal lies in their simplicity, low computational cost, and direct interpretability. 
\change{When these methods are applied to geometric representations, however, the resulting reduced coordinates primarily reconstruct geometric descriptors and do not generally provide a direct mapping back to the original design parameters. Recovering the native parameterization then requires an additional backmapping strategy.
Moreover, these methods remain intrinsically limited to linear subspaces of the original data space.}
This becomes restrictive whenever the sampled design family is governed by curved, folded, or otherwise nonlinear geometric variability, as frequently occurs in practical CAD-based and free-form parameterizations. In such cases, a linear reduced basis may require an unnecessarily large number of modes to achieve a prescribed reconstruction accuracy.

\change{Complementary dimensionality-reduction strategies operate directly in the original parameter space. Active-subspace methods \cite{lukaczyk2014active}, for example, identify low-dimensional directions along which selected quantities of interest vary most strongly, typically through gradient-based information. Because the reduction acts directly on the input parameters, the relation with the original design variables is retained. However, the resulting subspace is response-dependent and its construction generally requires function and gradient information, whereas geometry-driven dimensionality-reduction approaches can be constructed offline from the sampled design space, independently of a specific optimization objective or high-fidelity response.}

\change{More generally, the notion of a design manifold has been introduced to distinguish the nominal dimension of a parameterization from the intrinsic complexity of the associated family of designs \cite{chen2017design}. From this perspective,} nonlinear manifold-learning techniques \cite{d2017nonlinear,boncoraglio2021active,liu2026design,yan2026aerodynamic}, including autoencoders and related representation-learning methods \cite{bengio2013representation}, provide a natural \change{means of representing design families whose variability is concentrated near a curved low-dimensional manifold. Recent engineering applications have demonstrated the potential of learned nonlinear representations for constructing compact engineering design spaces, including simulation-ready autoencoder-based hull representations \cite{abbas2023deepmorpher} and controllable latent representations aligned with measurable vehicle attributes \cite{shintani2026controllable}.}
Yet, in engineering shape design, this additional expressivity comes with a major practical obstacle: the loss of a stable and operationally meaningful backmapping to the original parametric description. In simulation-driven workflows, reduced coordinates are useful only if they can be mapped back to feasible designs expressed in the same variables used to generate, modify, mesh, and analyze the geometry. Generic nonlinear autoencoding pipelines may reconstruct geometry accurately, but \change{a controlled parameter-mediated reconstruction through the original design variables is not an inherent part of their formulation}. This \emph{pre-image} limitation \cite{gaudrie2020modeling} is one of the main reasons why many nonlinear reduction methods remain difficult to deploy in CAD/CAE-driven design environments. 

Parametric Model Embedding (PME) \cite{serani2023parametric} was introduced to address the \change{native-parameter} backmapping limitation of geometry-based dimensionality reduction in parametric shape design. PME constructs reduced coordinates from geometric variability while preserving an explicit \change{analytical} link to the original design parameters. 
It does so by augmenting geometric data with the parameter vector itself and formulating dimensionality reduction as a weighted generalized principal component analysis in an augmented feature space. As a result, the latent representation is geometry-driven, whereas reconstruction remains analytically mediated by the original parametric variables. This property makes PME particularly suited to reduced-space design workflows in simulation-based engineering \cite{serani2024aerodynamic}. More recent developments have extended PME toward physics-informed and physics-driven formulations, showing that the framework can incorporate information beyond pure geometry while maintaining parametric consistency \cite{serani2025extending}. Nevertheless, the current PME family remains fundamentally linear in its latent structure.

This paper addresses that limitation by proposing a nonlinear extension of PME, referred to as nonlinear PME (NLPME). The objective is not to replace PME with a generic neural-network autoencoder, but to preserve its defining structural principle while extending its representational power beyond linear subspaces. The proposed framework retains the two features that make PME useful for engineering design: latent variables are inferred from geometric information, and reconstruction is constrained to pass through the original parametric design space. The nonlinear extension is therefore formulated as a structured latent representation in which the discretized geometry is encoded into a compact nonlinear latent variable, decoded into normalized design parameters, and finally mapped back to geometry through a forward parametric operator. In its ideal form, this operator coincides with the exact parametric generator. In the present proof-of-concept, it is implemented through a differentiable surrogate of the parameter-to-geometry map, enabling end-to-end training through a geometry-consistency objective.

%This distinction is essential. 
The proposed method should not be interpreted as a conventional geometric autoencoder, since geometry is not reconstructed directly from the latent variables. Instead, the latent coordinates are useful insofar as they generate admissible parameter vectors that remain compatible with the original design model. In this sense, the contribution of the paper is not simply the adoption of nonlinear mappings for dimensionality reduction, but the construction of a nonlinear reduced-dimensional representation that remains explicitly embedded in the original parametric design framework.

Structured autoencoder architectures that couple a forward physical model to the reconstruction pathway have been proposed in adjacent areas of computational science. In inverse-problem and surrogate-modeling settings, for instance, an encoder may compress observations into latent variables, a decoder may map those latent variables to physical parameters, and a forward operator may then enforce consistency with the original observations \cite{hart2026paired}. Similarly, in aerodynamic inverse design, variational encoder--decoder models operating over design-parameter vectors have been used to generate design portfolios consistent with prescribed performance targets \cite{Yang2023inversedesign}. These approaches share with the present work the idea of interposing a parameter space between the latent representation and the reconstructed output. However, they serve different objectives: they are primarily concerned with inverse recovery or generative design, whereas the present work addresses offline dimensionality reduction of an already defined parametric design space while preserving explicit compatibility with the original parametric variables.

The proposed methodology is assessed on the parametric design space of a bio-inspired autonomous underwater glider, characterized by a 32-dimensional geometric parameterization and a CAD-based geometry-generation process \cite{serani2026machine}. This test case is used as a methodological proof-of-concept rather than as a full design-optimization study. The dataset is interpreted as a structured Sobol discretization of the admissible design domain, and the quality of the embedding is assessed in terms of reconstruction accuracy, compression efficiency, and consistency of the parameter-mediated backmapping over the sampled design manifold. A deep autoencoder (DAE) baseline is also included to distinguish the effect of nonlinear compression from the effect of parameter-mediated reconstruction. \change{Accordingly, linear PME and DAE are used as complementary numerical baselines: the former isolates the effect of introducing nonlinear compression while retaining parameter-mediated reconstruction, whereas the latter provides a reference for nonlinear compression without this structural constraint.}

The main contributions of the paper are fourfold. First, a structured nonlinear extension of PME is introduced for dimensionality reduction in parametric shape design. Second, the role of a differentiable forward parametric operator is clarified by distinguishing between the ideal formulation based on the exact generator and the practical surrogate-based implementation adopted here. Third, a common evaluation framework is established to compare linear PME, NLPME, and DAE on the basis of a shared geometric reconstruction metric. Fourth, the proposed method is shown, on a representative parametric shape family, to improve significantly upon linear PME in terms of compression efficiency while retaining an explicit backmapping to the original design variables.

The remainder of the paper is organized as follows. Section~2 introduces the problem setting and notation. Section~3 briefly recalls the linear PME formulation. Section~4 presents the proposed nonlinear extension, including the structured architecture, the forward operator, the surrogate pre-training stage, and the training objective. Section~5 introduces the bio-inspired glider test case. Section~6 defines the evaluation strategy and metrics. Section~7 reports the numerical results. Section~8 discusses the implications and limitations of the proposed approach, and Section~9 summarizes the main conclusions.

\section{Problem setting and notation}
The present work addresses dimensionality reduction in the context of \emph{parametric shape optimization}, where admissible geometries are generated by a finite-dimensional design vector. In contrast with geometric shape optimization in the classical infinite-dimensional shape-calculus sense, the optimization variable is not the boundary itself, but a vector of design parameters belonging to a prescribed admissible set. Accordingly, let
\begin{equation}
\mathcal{U} \subset \mathbb{R}^{M}
\end{equation}
denote the admissible parametric design space, and let
\begin{equation}
\mathbf{u} \in \mathcal{U}
\end{equation}
be the vector of original design variables.

\begin{figure}[!t]
    \centering
    \includegraphics[width=0.75\linewidth]{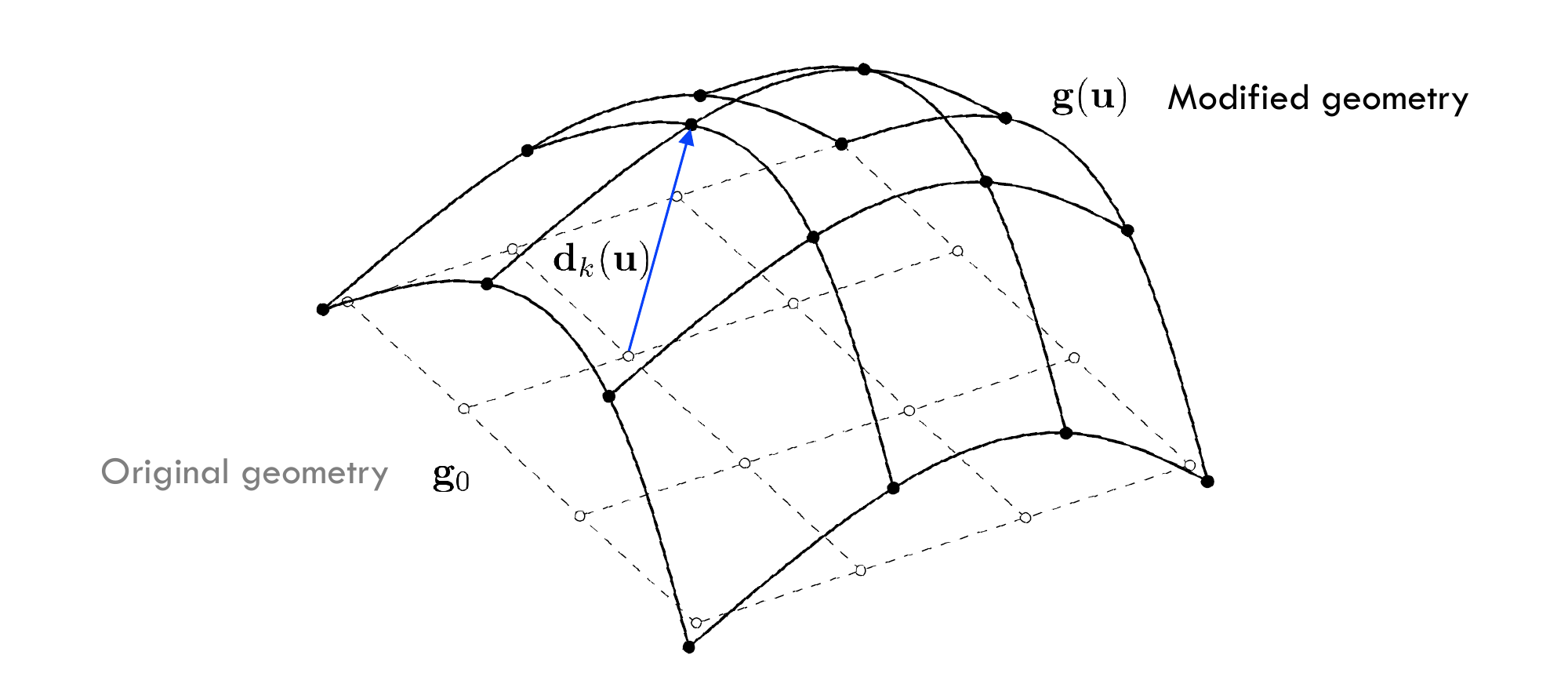}
    \caption{Schematic representation of the baseline geometry, the modified geometry, and the corresponding discrete shape-modification vector.}\label{fig:sketch}
\end{figure}

Each admissible parameter vector $\mathbf{u}$ defines a corresponding geometry. If needed, this geometry may be denoted by its continuous representation $\Omega(\mathbf{u})$; however, since the present work is entirely concerned with discrete geometric representations, the main object of interest is its discretized counterpart
\begin{equation}
\mathbf{g}(\mathbf{u}) \in \mathbb{R}^{n_g},
\end{equation}
where $n_g$ denotes the number of geometric degrees of freedom used to represent the shape. Depending on the application, $\mathbf{g}(\mathbf{u})$ may contain, for instance, nodal coordinates, surface point coordinates, or any other aligned geometric descriptor (see Fig. \ref{fig:sketch}). \change{The formulation is not restricted to three-dimensional geometries, and can in principle be applied to parametric configurations of different dimensionality, provided that a consistent discrete representation and a corresponding parametric mapping are available.}
Introducing the forward parametric-to-geometry operator
\begin{equation}
\mathcal{G} : \mathcal{U} \rightarrow \mathbb{R}^{n_g},
\qquad
\mathbf{g}(\mathbf{u}) = \mathcal{G}(\mathbf{u}),
\end{equation}
let $\mathbf{g}_0 \in \mathbb{R}^{n_g}$ denote a baseline or reference geometry. Following the standard notation of parametric model embedding, the associated shape modification vector is defined as
\begin{equation}
\mathbf{d}(\mathbf{u}) = \mathbf{g}(\mathbf{u}) - \mathbf{g}_0.
\end{equation}
Thus, the original parametric design pipeline may be written as
\begin{equation}
\mathbf{u} \in \mathcal{U}
\quad \xrightarrow{\;\mathcal{G}\;}
\quad
\mathbf{g}(\mathbf{u})
\quad \longrightarrow \quad
\mathbf{d}(\mathbf{u}) = \mathbf{g}(\mathbf{u}) - \mathbf{g}_0.
\end{equation}

In a general parametric shape optimization setting, one may seek the solution of a problem of the form
\begin{equation}
\min_{\mathbf{u} \in \mathcal{U}} \mathcal{J}\big(\mathbf{g}(\mathbf{u})\big),
\end{equation}
or, more generally,
\begin{equation}
\min_{\mathbf{u} \in \mathcal{U}} \widetilde{\mathcal{J}}\big(\mathbf{y}(\mathbf{u}),\mathbf{g}(\mathbf{u})\big),
\end{equation}
where $\mathbf{y}(\mathbf{u})$ denotes a state variable, a physical response, or any quantity of interest associated with the geometry generated by $\mathbf{u}$. However, the present paper does not address the optimization problem itself. Instead, it focuses on the preliminary and independent task of constructing a reduced-dimensional representation of the sampled design space, suitable for future use in optimization, surrogate modeling, or design-space exploration.
\change{Accordingly, the dimensionality reduction considered here acts on the parametric design space and should be distinguished from model-order reduction of the state problem. NLPME does not directly reduce the number of degrees of freedom or the size of the algebraic systems arising from the discretization of governing equations, although the resulting lower-dimensional design representation may reduce the overall computational effort of downstream design-space exploration, optimization, surrogate modeling, or uncertainty quantification.}

Let $\mathbf{u}^{(i)} \in \mathcal{U}$, $i=1,\dots,S$, denote a set of $S$ sampled admissible parameter vectors. For each sample,
\begin{equation}
\mathbf{g}^{(i)} = \mathcal{G}\big(\mathbf{u}^{(i)}\big),
\qquad
\mathbf{d}^{(i)} = \mathbf{g}^{(i)} - \mathbf{g}_0.
\end{equation}
%In the present work, this dataset is generated by a Sobol sampling strategy and is therefore interpreted as a structured discretization of the design space, rather than as a random sample drawn from an unknown population. 
The dataset is generated by Sobol sampling and is interpreted as a structured discretization of the design space.
This interpretation is important because the goal is not to learn a predictor valid beyond the sampled domain, but to construct a compact representation of the sampled design manifold itself.

The reduced representation problem considered in this work can then be stated as follows. Given the sampled dataset, identify a latent variable vector
\begin{equation}
\mathbf{z} \in \mathbb{R}^{N},
\qquad N \ll M,
\end{equation}
together with a nonlinear encoding map
\begin{equation}
\mathcal{E} : \mathbb{R}^{n_g} \rightarrow \mathbb{R}^{N},
\qquad
\mathbf{z} = \mathcal{E}(\mathbf{d}),
\end{equation}
and a decoding map
\begin{equation}
\mathcal{D} : \mathbb{R}^{N} \rightarrow \mathbb{R}^{M},
\qquad
\hat{\mathbf{u}} = \mathcal{D}(\mathbf{z}),
\end{equation}
such that the latent variable captures the dominant structure of the sampled design space while preserving a consistent reconstruction through the original parametric description.

This requirement distinguishes the present formulation from purely geometric dimensionality reduction. In standard geometry-based reduction, one typically seeks a compressed representation allowing the direct reconstruction of $\mathbf{g}$ or $\mathbf{d}$ in the same geometric space. Here, instead, the reconstruction is constrained to occur through the parametric variables themselves. In its ideal form, the reconstructed parameter vector $\hat{\mathbf{u}}$ is first mapped back to the geometry through the original forward operator, yielding
\begin{equation}\label{eq:goper}
\hat{\mathbf{g}} = \mathcal{G}\big(\hat{\mathbf{u}}\big),
\qquad
\hat{\mathbf{d}} = \hat{\mathbf{g}} - \mathbf{g}_0.
\end{equation}

In the following, this parameter-mediated reconstruction requirement is used as the structural constraint for the nonlinear extension of PME.

\section{Linear parametric model embedding}
Before introducing the proposed nonlinear extension, we recall the
essential ingredients of linear PME that motivate its nonlinear
generalization \cite{serani2023parametric}.

Using the notation introduced in Section~2, let $\mathbf{D}\in\mathbb{R}^{n_g\times S}$ and $\mathbf{U}\in\mathbb{R}^{M\times S}$ denote the matrices collecting, column-wise, the sampled shape-modification vectors and the corresponding design variables, respectively:
\begin{equation}
\mathbf{D} =
\left[
\mathbf{d}^{(1)} \ \cdots \ \mathbf{d}^{(S)}
\right]
\in \mathbb{R}^{n_g \times S},
\end{equation}
\begin{equation}
\mathbf{U} =
\left[
\mathbf{u}^{(1)} \ \cdots \ \mathbf{u}^{(S)}
\right]
\in \mathbb{R}^{M \times S}.
\end{equation}

Classical geometry-based PCA constructs a reduced representation by analyzing only the matrix $\mathbf{D}$, thereby identifying directions of dominant geometric variance. Although this yields a compact linear description of the sampled shape space, it does not preserve an explicit mapping back to the original design variables. This is precisely the pre-image problem that PME was introduced to address.

The basic idea of PME is to augment the geometric data with the original design variables, leading to the generalized feature matrix
\begin{equation}
\mathbf{P} =
\begin{bmatrix}
\mathbf{D}\\
\mathbf{U}
\end{bmatrix}
\in \mathbb{R}^{(n_g+M)\times S}.
\end{equation}
%The dimensionality-reduction problem is then formulated as a generalized principal component analysis in the augmented feature space. 
The dimensionality-reduction problem is then formulated as a generalized principal component analysis in \change{a weighted inner-product space, corresponding to the discrete formulation of the Karhunen--Lo\`eve expansion \cite{diez2015-CMAME}, and extended to the augmented PME feature space \cite{serani2023parametric}}.
In its most general form, PME solves the eigenvalue problem
\begin{equation}
\mathbf{A}\mathbf{G}\mathbf{W}\mathbf{Z}
=
\mathbf{Z}\boldsymbol{\Lambda},
\qquad
\mathbf{A}=\frac{1}{S}\mathbf{P}\mathbf{P}^{T},
\end{equation}
\change{where $\mathbf G$ and $\mathbf W$ are diagonal matrices defined over the augmented feature space, $\mathbf Z$ is the matrix of generalized eigenvectors, and $\boldsymbol\Lambda$ is the diagonal matrix of generalized eigenvalues. The metric matrix has the block structure $\mathbf G=\operatorname{diag}(\mathbf G_d,\mathbf I),$ where $\mathbf G_d$ accounts for the measures associated with the discrete geometric elements and $\mathbf I$ is the identity matrix in the parameter space. The weighting matrix has the block structure $\mathbf W=\operatorname{diag}(\mathbf W_d,\mathbf 0),$ where $\mathbf W_d$ defines the weighting of the geometric information, while the null parametric block assigns zero weight to the original design variables.}

This formulation is a weighted generalized PCA, rather than a singular value decomposition, and this distinction is essential because it allows explicit control of the contribution associated with the geometry and parameters \cite{serani2025extending}.

\change{The null parametric block in $\mathbf W$ defines the \emph{zero-weighting paradigm} of PME. Consequently, the latent space is constructed exclusively from geometric variability, while the original design variables remain embedded in the augmented feature space. The reduced coordinates are therefore geometry-driven, whereas the parametric components of the associated generalized eigenvectors provide the information required to reconstruct the original parameterization.}

Let $N$ denote the reduced dimensionality selected according to a prescribed retained-information criterion. Then, the matrix of the first $N$ normalized generalized eigenvectors can be partitioned as
\begin{equation}
\mathbf{Z}_N =
\begin{bmatrix}
\mathbf{Q}_N\\
\mathbf{V}_N
\end{bmatrix},
\end{equation}
where $\mathbf{Q}_N$ contains the geometric components and $\mathbf{V}_N$ the parametric components of the reduced basis. Let $\boldsymbol{\alpha}\in\mathbb{R}^{N}$ denote the reduced coordinates associated with a given sampled configuration. In the standard centered formulation, the parametric backmapping is written as
\begin{equation}
\hat{\mathbf{u}}
=
\bar{\mathbf{u}}
+
\mathbf{V}_N \boldsymbol{\alpha},
\label{eq:pme_backmapping}
\end{equation}
where $\bar{\mathbf{u}}$ is the empirical mean of the sampled parameter vectors. PME therefore provides a reduced representation that is simultaneously geometry-driven in the construction of the latent space and analytically invertible toward the original parametric variables. \change{A publicly available reference implementation of linear PME is provided
by PME-toolkit \cite{serani_2026_19068340}.}

%This property distinguishes PME from standard PCA applied to geometry alone. In conventional geometry-based reduction, the latent variables reconstruct only geometric descriptors, and an additional pre-image strategy is required to recover the original design parameters, if such recovery is possible at all. In PME, instead, the original parameters are part of the generalized feature space from the outset, and their reconstruction follows directly from the reduced representation.

%The resulting reduced-dimensional model may therefore be summarized by the linear chain $\mathbf{d} \;\mapsto\; \boldsymbol{\alpha} \;\mapsto\; \hat{\mathbf{u}}$. This structure resolves the main backmapping limitation of geometry-only PCA and makes PME particularly suited to design-space dimensionality reduction in parametric shape-optimization workflows.

\change{This property distinguishes PME from standard geometry-based PCA, where an additional pre-image strategy is required to recover the original design parameters. In PME, the parameters are embedded in the generalized feature space and their reconstruction follows directly from the reduced representation, which may be summarized by the linear chain $\mathbf{d}
\;\mapsto\;
\boldsymbol{\alpha}
\;\mapsto\;
\hat{\mathbf{u}}$.} At the same time, the reduced manifold represented by PME remains linear, since both the encoding and the reconstruction are based on linear projections in the augmented feature space. Consequently, whenever the sampled design space exhibits intrinsically nonlinear geometric structure, the linear subspace identified by PME may require an unnecessarily large number of reduced variables to achieve a given reconstruction accuracy. This limitation motivates the nonlinear extension proposed in the next section.

\section{Nonlinear parametric model embedding}
\label{sec:nlpme}

The proposed nonlinear extension is built on the structural principle underlying PME: reduced coordinates are inferred from geometric variability, while reconstruction remains explicitly mediated by the original parametric description. The objective is not to replace PME with a generic nonlinear autoencoder, but to preserve its geometry-driven and parameter-mediated logic while relaxing the linear latent-space assumption. Accordingly, NLPME replaces the linear projection--reconstruction mechanism of PME with nonlinear mappings, while retaining the requirement that latent variables first reconstruct the original design parameters and only then recover the corresponding geometry.

The nonlinear latent representation is defined as
\begin{equation}
\mathbf{z} \in \mathbb{R}^{N},
\qquad N \ll M,
\end{equation}
where \(N\) is the reduced dimension and \(M\) is the dimension of the original parametric design space. %Unlike standard nonlinear autoencoding approaches, the goal is not to reconstruct geometry directly from \(\mathbf{z}\), but to reconstruct the original design parameters first and then recover the corresponding geometry through a forward parametric map. The resulting formulation is therefore a structured nonlinear embedding rather than a generic encoder--decoder operating entirely in geometry space.

\subsection{Structured architecture}

The proposed nonlinear PME (NLPME) is based on the following chain:
\begin{equation}
\mathbf{d}
\;\xrightarrow{\;\mathcal{E}_\phi\;}\;
\mathbf{z}
\;\xrightarrow{\;\mathcal{D}_\theta\;}\;
\hat{\mathbf{u}}
\;\xrightarrow{\;\mathcal{G}\ \text{or}\ \mathcal{S}_{\psi^*}\;}\;
\hat{\mathbf{g}}
\;\longrightarrow\;
\hat{\mathbf{d}} = \hat{\mathbf{g}} - \mathbf{g}_0,
\label{eq:nlpme_chain_updated}
\end{equation}
where:
\begin{itemize}
    \item $\mathcal{E}_\phi$ is a nonlinear encoder, parameterized by weights $\phi$, mapping the input geometric representation to the latent space;
    \item $\mathcal{D}_\theta$ is a nonlinear decoder, parameterized by weights $\theta$, mapping the latent variables to reconstructed design parameters;
    \item $\mathcal{G}$ denotes the exact forward parametric generator, when available;
    \item $\mathcal{S}_{\psi^*}$ denotes a differentiable surrogate approximation of the forward parametric map, pre-trained and frozen during NLPME training.
\end{itemize}

In functional form, the nonlinear embedding can be written as
\begin{equation}
\mathbf{z} = \mathcal{E}_\phi(\mathbf{d}),
\qquad
\hat{\mathbf{u}} = \mathcal{D}_\theta(\mathbf{z}),
\qquad
\hat{\mathbf{d}} = \mathcal{S}_{\psi^*}(\hat{\mathbf{u}}),
\label{eq:functional_chain}
\end{equation}
or, in the ideal case where the exact generator is directly available in differentiable form,
\begin{equation}
\hat{\mathbf{d}} = \mathcal{G}(\hat{\mathbf{u}})-\mathbf{g}_0.
\label{eq:functional_chain_exact}
\end{equation}

\begin{figure}[!t]
    \centering
    \includegraphics[width=1\linewidth]{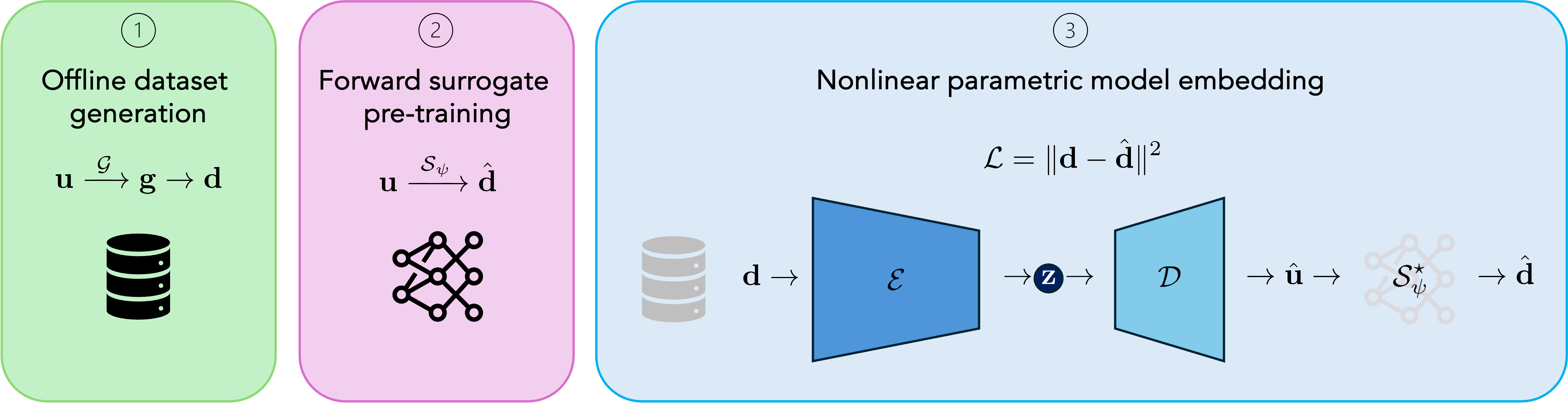}
\caption{Schematic representation of the proposed NLPME workflow. \textbf{Phase 1}: offline generation of admissible parametric designs and corresponding discrete shape modifications. \textbf{Phase 2}: supervised pre-training of a differentiable surrogate of the forward parametric map, subsequently frozen. \textbf{Phase 3}: training of the nonlinear embedding by encoding the shape modification vector into a latent variable, decoding it into design parameters, and reconstructing geometry through the frozen surrogate under a geometry-consistency loss.}
    \label{fig:scheme}
\end{figure}

The crucial feature of \eqref{eq:nlpme_chain_updated} is that the reconstruction is not performed directly as $\hat{\mathbf{d}}=\mathcal{R}(\mathbf{z})$ for some generic geometric decoder $\mathcal{R}$. Instead, the latent variable first reconstructs a parameter vector, and geometry is then recovered through the corresponding forward map. This makes the latent representation operationally meaningful from a design perspective, because every reconstructed configuration remains expressed in terms of admissible parametric variables.
A schematic overview of the full workflow is reported in Fig.~\ref{fig:scheme}, while the corresponding training procedure is summarized in Algorithm~\ref{alg:nlpme}.

The encoder receives only geometric information as input. In the current formulation, this input is the preprocessed discrete shape-modification vector associated with each sampled geometry. Therefore, the latent space is constructed from geometric variability alone, consistently with the zero-weighting logic of linear PME, where the original parameters do not contribute directly to the reduced coordinates. At the same time, because the decoder reconstructs the original design variables, the resulting latent representation remains tied to the parametric model. NLPME is therefore not an exact nonlinear analogue of the generalized eigenvalue problem underlying PME, but a structured nonlinear extension of the same design principle: the reduced coordinates are geometry-driven, while the reconstruction remains parameter-mediated.

\subsection{Forward parametric operator: exact and surrogate forms}
Consistently with the parameter-mediated reconstruction introduced in Eq.~\eqref{eq:goper}, the ideal form of NLPME reconstructs the geometry through the original parametric generator:
\begin{equation}
\hat{\mathbf{g}} = \mathcal{G}(\hat{\mathbf{u}}),
\qquad
\hat{\mathbf{d}} = \hat{\mathbf{g}} - \mathbf{g}_0.
\label{eq:exact_forward_updated}
\end{equation}
This formulation is the most faithful to the original PME philosophy, since it preserves the direct connection with the underlying parametric design model.

In practice, however, the exact operator $\mathcal{G}$ may not be directly usable within a gradient-based training loop. This may happen for several reasons: the original parameterization tool may be external to the learning environment, non-differentiable, computationally expensive, or unavailable when working from precomputed datasets. For this reason, in the present implementation, $\mathcal{G}$ is replaced with a differentiable surrogate operator $\mathcal{S}_{\psi^*}$, obtaining
\begin{equation}
\hat{\mathbf{g}} = \mathcal{S}_{\psi^*}(\hat{\mathbf{u}}),
\qquad
\hat{\mathbf{d}} = \hat{\mathbf{g}} - \mathbf{g}_0.
\label{eq:surrogate_forward_updated}
\end{equation}

%It is important to stress that the surrogate is not part of the conceptual definition of NLPME. Rather, it is an implementation-level proxy of the original forward operator. Its role is twofold. First, it provides a computationally efficient approximation of the parametric map. Second, and more importantly, it enables end-to-end differentiable training through a geometry-consistency objective. In principle, whenever a differentiable and sufficiently efficient exact parameterization is available, the surrogate can be removed and \eqref{eq:exact_forward_updated} can be used directly.

The surrogate is used here only as a differentiable implementation of the forward parametric map. Whenever the exact generator is differentiable and sufficiently efficient, Eq.~\eqref{eq:exact_forward_updated}  can be used directly.

\subsection{Surrogate pre-training}

The surrogate operator introduced above is trained in a supervised
offline stage before learning the nonlinear embedding. Let
\(\mathcal S_\psi\) denote a neural approximation of the forward
parametric map, parameterized by weights \(\psi\), such that
\begin{equation}
\mathcal{S}_{\psi} : \mathbb{R}^{M} \rightarrow \mathbb{R}^{n_g}.
\end{equation}
The surrogate is trained on the offline database of sampled admissible configurations by minimizing the discrepancy between the target geometric representation and the surrogate prediction:
\begin{equation}
\psi^* =
\operatorname*{argmin}_{\psi}
\frac{1}{S}
\sum_{i=1}^{S}
\left\|
\mathbf{d}^{(i)} - \mathcal{S}_{\psi}\!\left(\mathbf{u}^{(i)}\right)
\right\|^2.
\label{eq:surrogate_loss_updated}
\end{equation}
After convergence, the surrogate parameters are frozen and the trained model $\mathcal{S}_{\psi^*}$ is used as a differentiable proxy of the exact forward operator throughout the NLPME training stage.

Once frozen, \(\mathcal{S}_{\psi^\ast}\) enables gradients of the geometry-consistency loss to be propagated back to the decoder and encoder through the chain rule. This surrogate-learning stage corresponds to Phase 2 in Fig.~\ref{fig:scheme} and Algorithm~\ref{alg:nlpme}.

\begin{algorithm}[!t]
\caption{NLPME workflow}
\label{alg:nlpme}
\begin{algorithmic}[1]
\Require Sampled dataset $\{(\mathbf{u}^{(i)},\mathbf{d}^{(i)})\}_{i=1}^{S}$, latent dimension $N$
\Ensure Trained encoder $\mathcal{E}_{\phi^*}$, decoder $\mathcal{D}_{\theta^*}$, and latent representation $\mathbf{z}$

\Statex \textbf{Phase 1: Offline dataset generation}
\State Construct the sampled dataset of admissible parametric designs and corresponding shape modifications

\Statex \textbf{Phase 2: Forward surrogate pre-training}
\State Pre-train the surrogate $\mathcal{S}_{\psi}$ by minimizing the surrogate loss in Eq.~\eqref{eq:surrogate_loss_updated}
\State Freeze the surrogate parameters and retain $\mathcal{S}_{\psi^*}$

\Statex \textbf{Phase 3: NLPME training}
\State Initialize encoder parameters $\phi$ and decoder parameters $\theta$
\Repeat
    \For{$i=1,\dots,S$}
        \State $\mathbf{z}^{(i)}=\mathcal{E}_{\phi}(\mathbf{d}^{(i)})$
        \State $\hat{\mathbf{u}}^{(i)}=\mathcal{D}_{\theta}(\mathbf{z}^{(i)})$
        \State $\hat{\mathbf{d}}^{(i)}=\mathcal{S}_{\psi^*}(\hat{\mathbf{u}}^{(i)})$
    \EndFor
    \State Update $\phi,\theta$ by minimizing the geometry-consistency loss in Eq.~\eqref{eq:geom_loss_updated}
\Until{convergence}
\State \Return $\mathcal{E}_{\phi^*}, \mathcal{D}_{\theta^*}$ and the corresponding latent variables
\end{algorithmic}
\end{algorithm}

\subsection{Training objective}
The model is trained exclusively through a geometry-consistency objective. \change{To retain the general metric formulation introduced for linear PME, let $\mathbf G_d\in\mathbb R^{n_g\times n_g}$ denote the geometric block of the augmented metric matrix $\mathbf G$. The matrix $\mathbf G_d$ defines the inner product associated with the discrete geometric representation and may account, for instance, for the measures associated with a nonuniform geometric discretization. In general, the geometry-consistency loss is therefore defined as}
\begin{equation}
\change{
\mathcal L_{\mathrm{geom}}
=
\frac{1}{S}
\sum_{i=1}^{S}
\left(
\mathbf d^{(i)}-\hat{\mathbf d}^{(i)}
\right)^{T}
\mathbf G_d
\left(
\mathbf d^{(i)}-\hat{\mathbf d}^{(i)}
\right).
}
\label{eq:geom_loss_general}
\end{equation}
\change{For the standardized geometric coordinates used in the present neural implementation, the loss is evaluated using the Euclidean inner product, corresponding to $\mathbf G_d=\mathbf I$. Consequently, Eq.~\eqref{eq:geom_loss_general} reduces exactly to}
\begin{equation}
\mathcal L_{\mathrm{geom}}
=
\frac{1}{S}
\sum_{i=1}^{S}
\left\|
\mathbf d^{(i)}-\hat{\mathbf d}^{(i)}
\right\|^2.
\label{eq:geom_loss_updated}
\end{equation}
\change{Equation~\eqref{eq:geom_loss_updated} is the geometry-consistency loss used in the numerical implementation.} 
Thus, the latent representation is learned by requiring that the decoded parameters, once passed through the forward parametric operator, reconstruct the original geometric modifications as accurately as possible.

No explicit parameter-reconstruction term is included in the present formulation. Likewise, no additional boundary penalty is introduced in the loss. In the actual implementation, the decoder output is constrained through a sigmoid activation to remain within the normalized admissible range of the design parameters. To avoid unnecessary notation overload, however, the methodological formulation is written directly in terms of the parameter vector $\mathbf{u}$.

The exclusive use of \eqref{eq:geom_loss_updated} is intentional. The purpose of the present paper is to assess whether a geometry-driven nonlinear latent representation, constrained to reconstruct through the original parametric description, can improve compression efficiency with respect to linear PME. More general loss formulations involving, for instance, direct parameter-consistency terms or additional regularization components, are possible and may be explored in future developments, but they are not required for the present methodological assessment.
The overall NLPME training stage, including the encoder--decoder reconstruction chain through the frozen surrogate, corresponds to Phase~3 in Fig.~\ref{fig:scheme} and Algorithm~\ref{alg:nlpme}.

\section{Test case: bio-inspired glider design space}
\label{sec:test_case}

The proposed nonlinear extension is assessed on the parametric design space of a bio-inspired autonomous underwater glider. This test case is well suited to the present proof-of-concept because it combines a relatively high-dimensional parameterization with a CAD-based geometry-generation pipeline, yielding a family of admissible shapes that is rich enough to challenge linear dimensionality reduction while remaining embedded in a consistent parametric design framework.

\begin{figure}[!t]
    \centering
    \includegraphics[width=1\linewidth]{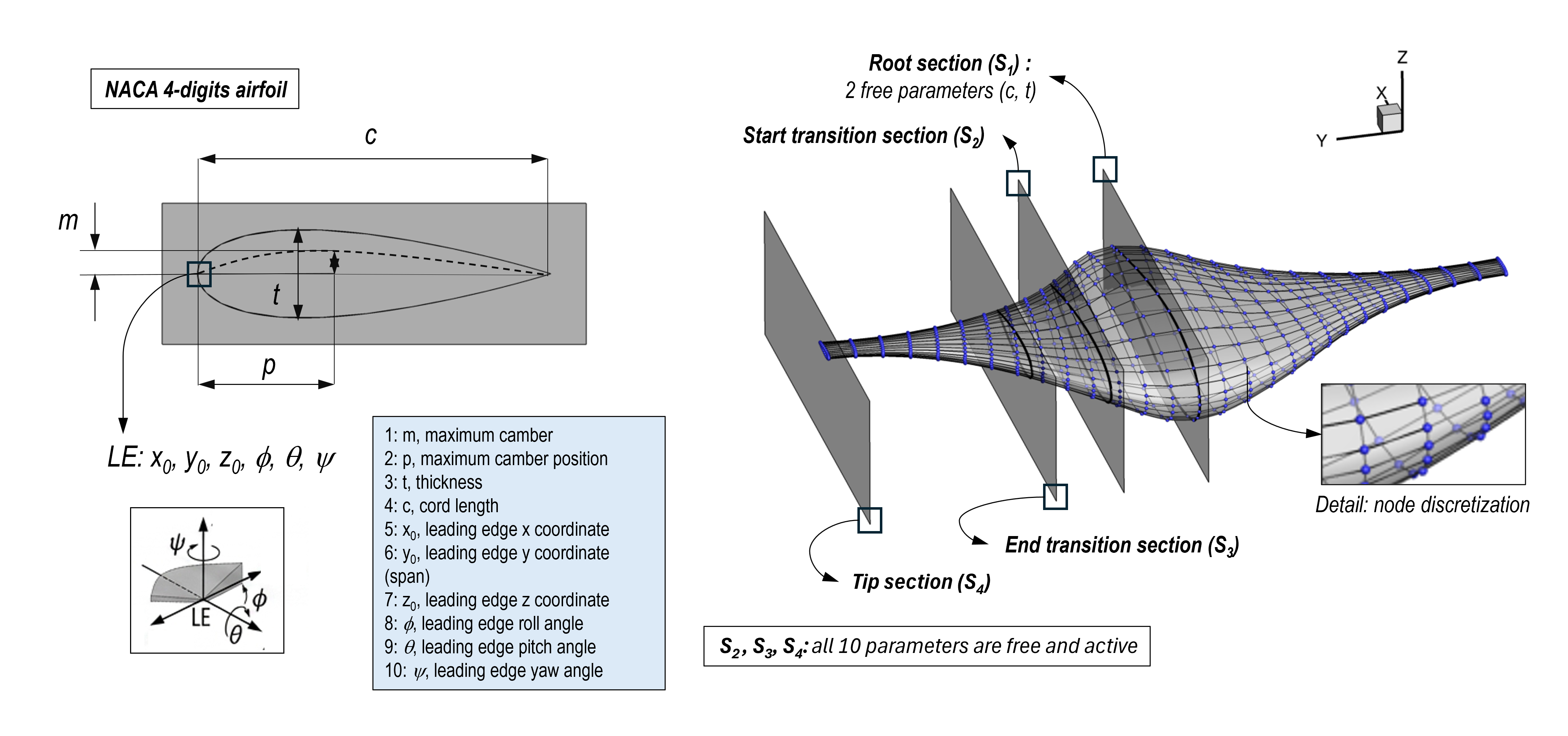}
    \caption{Parametric model for the bio-inspired autonomous underwater glider. Left: NACA 4-digit-based section parameterization with leading-edge coordinates and orientation angles. Right: three-dimensional geometry and discrete surface representation used for dimensionality reduction.}
    \label{fig:param}
\end{figure}

The external shell is inspired by the morphology of a manta ray and is constructed from four transverse sections distributed from the central body to the wing tip (see Fig.~\ref{fig:param}). Each section is defined through a NACA-4-digit-based airfoil description combined with spatial transformations. More precisely, each section includes four airfoil parameters---maximum camber, camber position, thickness ratio, and chord---together with the three coordinates of the leading-edge position and three angular rotations, namely twist, roll, and yaw~\cite{serani2025preliminary}.

Because of symmetry and continuity requirements at the root section, the total number of free geometric parameters for the half-body reduces to \(M=32\). In particular, only two independent variables are retained at the root section, while all ten parameters are active for each of the remaining three sections. This parameterization provides sufficient flexibility to alter the planform, local section shapes, sweep-like effects, spanwise twist, and three-dimensional wing-body blending, while preserving a compact and physically meaningful set of original design variables.

Although the number of design variables is finite, the resulting map from design parameters to geometry is not a simple linear superposition of independent shape modes. The four transformed sections are assembled through a CAD construction pipeline based \change{on the Open CASCADE Technology geometric modeling kernel
\cite{opencascade2026}}. In the reference implementation, each section is first built as a closed airfoil-like curve, and the outer shell is then generated through a lofting procedure across sections. Additional geometric treatments are introduced to ensure robustness and regularity of the final shell, including trailing-edge regularization, spanwise tangency constraints, repeated tip sections, and final trimming operations.

These operations induce nontrivial geometric couplings among the original variables. Local changes in sectional parameters may propagate nonlinearly through the lofted surface, while global modifications of the outer shell may result from coupled variations of section shape, position, and orientation. Therefore, even though the parameterization is finite-dimensional, the corresponding family of discretized geometries is expected to occupy a curved manifold in the geometric space rather than a simple low-dimensional linear subspace. This makes the test case relevant for assessing whether a nonlinear extension of PME can improve compression efficiency over its linear counterpart.

The dataset is generated by Sobol sampling of the original 32-dimensional geometric design space~\cite{serani_2026_18936594}. A total of \(S=16{,}385\) configurations are considered, consistently with the reference design-space exploration previously carried out for this geometry~\cite{serani2026machine}. Of these configurations, \(7{,}467\) valid geometries are retained after filtering~\cite{serani2025extending}. For each valid design, the corresponding outer shell is generated through the CAD pipeline and converted into a fixed discrete geometric representation suitable for dimensionality reduction. The resulting family of sampled shapes defines the design manifold on which both linear PME and the proposed NLPME are constructed.

%The numerical results reported in the following sections should therefore be interpreted as an assessment of representational capability, namely the ability of the reduced coordinates to describe this sampled design space compactly and consistently, rather than as the outcome of a downstream shape-optimization exercise.

\section{Evaluation strategy and metrics}
\label{sec:evaluation}

% A key aspect of the present formulation concerns the role of the dataset in the construction and assessment of the reduced representation. Unlike predictive machine-learning settings, the objective of NLPME is not to infer a model that generalizes to arbitrary unseen geometries outside the sampled design domain. Rather, the goal is to construct a compact nonlinear representation of a given parametric design space, sampled in a structured manner and treated as a discrete approximation of the admissible family of shapes.

% For this reason, the full set of valid sampled configurations is used to construct and evaluate the reduced representations. This choice is consistent with classical offline dimensionality-reduction methods, such as PCA, POD, and linear PME, where the reduced space is computed from the available samples describing the design manifold of interest. In the present setting, the purpose of the learning procedure is therefore not statistical prediction on an external test distribution, but geometric compression of the sampled design space while preserving parametric consistency.

The reduced representations are constructed and assessed in an offline dimensionality-reduction setting. The valid Sobol-sampled configurations are treated as a structured discretization of the admissible design space, rather than as random samples from an unknown population. Accordingly, the purpose of the learning procedure is geometric compression of the sampled design manifold while preserving parametric consistency, not prediction on an external test distribution.

The same principle is adopted for the comparison among linear PME, DAE, and NLPME. All methods operate on the same set of admissible geometries and are evaluated using the same geometric reconstruction metric. The only exception concerns the surrogate of the forward parametric map used inside NLPME: during surrogate pre-training, a small validation subset is used only for model selection of the surrogate itself. Once trained, the surrogate is frozen and the NLPME encoder--decoder is trained on the full valid dataset. This validation step is therefore instrumental to the differentiable implementation of the forward map and does not define a train/test evaluation protocol for the reduced representation.

The DAE baseline is also trained on the full valid dataset, consistently with its role as a nonlinear geometric compression baseline. However, because the DAE reconstructs directly in the geometric space and does not enforce reconstruction through the original parametric variables, a light \(L^2\) weight decay, equal to $2.5 \times10^{-4}$, is applied during training. 
%This regularization is used to mitigate unconstrained memorization in the direct geometric decoder. The same explicit weight decay is not applied to NLPME, whose reconstruction is structurally constrained to pass through the normalized design-variable space and the frozen forward surrogate.
\change{\change{This regularization provides mild capacity control for the DAE, whose direct geometric reconstruction pathway is not constrained to pass through the original design-variable space or a forward parametric operator. The same explicit weight decay is not applied to NLPME, whose reconstruction is structurally constrained to pass through the normalized design-variable space and the frozen forward surrogate.}}

To compare the reduced representations on a common basis, we adopt a shared geometric reconstruction \change{accuracy} metric. Let
\begin{equation}
\bar{\mathbf d}
=
\frac{1}{S}
\sum_{i=1}^{S}
\mathbf d^{(i)}
\end{equation}
denote the empirical mean of the sampled shape-modification vectors, and let \(\hat{\mathbf d}^{(i)}(N)\) denote the reconstruction of the \(i\)-th sample obtained from a reduced model of latent dimension \(N\). \change{The reconstruction accuracy is quantified through the geometric normalized mean squared error, defined as}
\begin{equation}
\epsilon(N)
=
\frac{
\displaystyle
\sum_{i=1}^{S}
\left\|
\mathbf d^{(i)}
-
\hat{\mathbf d}^{(i)}(N)
\right\|^2
}{
\displaystyle
\sum_{i=1}^{S}
\left\|
\mathbf d^{(i)}
-
\bar{\mathbf d}
\right\|^2
}.
\label{eq:nmse}
\end{equation}
Equation~\eqref{eq:nmse} is used throughout this work as the common performance measure for all reduced representations. In the case of linear PME, \(\hat{\mathbf d}^{(i)}(N)\) is obtained through projection onto the linear basis truncated to \(N\) components. In the case of NLPME, it is obtained through the nonlinear encoder--decoder chain and the corresponding forward geometric reconstruction. In the case of DAE, it is obtained directly from the geometric decoder.

In the linear setting, \(\epsilon(N)\) coincides with the residual fraction of geometric variance not captured by the retained subspace. In the nonlinear setting, no equivalent variance decomposition is generally available; nevertheless, Eq.~\eqref{eq:nmse} provides a direct and operationally consistent basis for comparing the compression capability of linear and nonlinear reduced models at fixed reconstruction accuracy.

In the numerical results, two practically relevant thresholds are considered: a target reconstruction level corresponding to \(\epsilon \leq 5\%\), and a more stringent level corresponding to \(\epsilon \leq 1\%\). 
Compression efficiency is then assessed by comparing the latent dimension required by each method to achieve the same geometric reconstruction threshold.

In addition to the global reconstruction error in Eq.~\eqref{eq:nmse}, per-sample normalized squared errors are used to analyze how the reconstruction error is distributed across the sampled design space. For each sample \(i\), this quantity is defined as
\begin{equation}
\epsilon_i(N)
=
\frac{
\left\|
\mathbf d^{(i)}
-
\hat{\mathbf d}^{(i)}(N)
\right\|^2
}{
\displaystyle
\frac{1}{S}
\sum_{j=1}^{S}
\left\|
\mathbf d^{(j)}
-
\bar{\mathbf d}
\right\|^2
}.
\label{eq:nse_sample}
\end{equation}
With this definition, the average of \(\epsilon_i(N)\) over the dataset is equal to the global NMSE \(\epsilon(N)\). The per-sample distribution therefore provides a sample-wise counterpart of the global reconstruction metric and is used to verify whether the observed compression gain is representative of the sampled design space rather than dominated by a small subset of
configurations.

\section{Results}
\label{sec:results}

\begin{figure}[!b]
    \centering
    \includegraphics[width=0.45\linewidth]{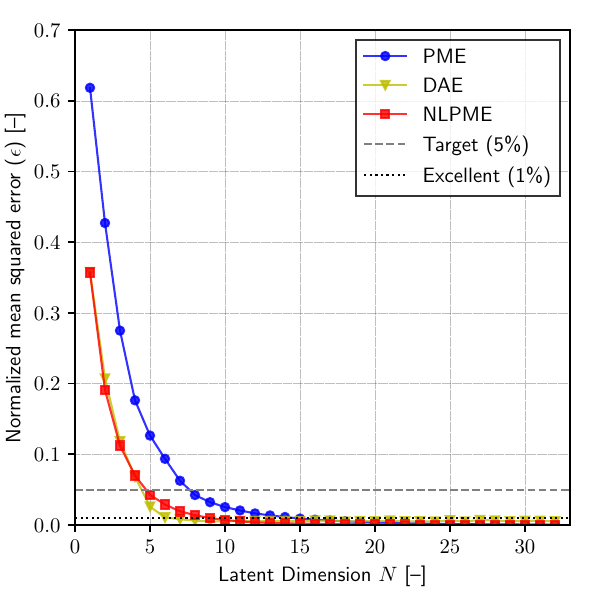}
    \includegraphics[width=0.45\linewidth]{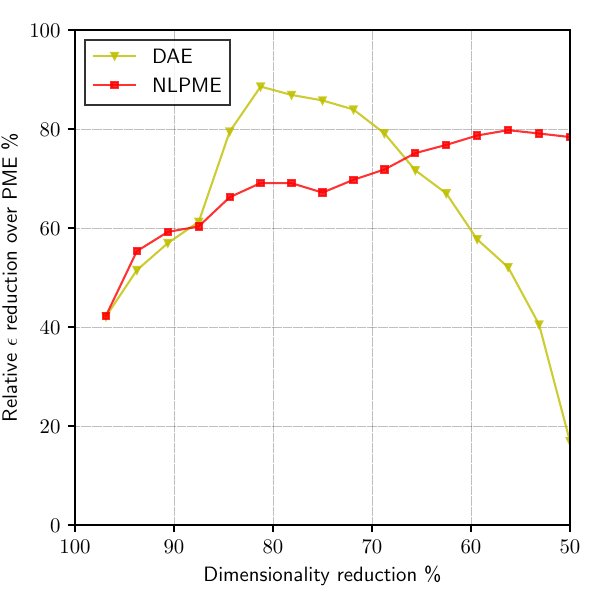}
    \caption{Geometric reconstruction performance of PME, DAE, and NLPME as a function of the latent dimension \(N\). Left: normalized mean squared reconstruction error \(\epsilon(N)\). The dashed and dotted horizontal lines indicate the 5\% target and 1\% stringent reconstruction levels, respectively. Right: relative reduction of \(\epsilon\) with respect to PME at equal latent dimension, plotted as a function of dimensionality reduction \(100(1-N/32)\).}
    \label{fig:nmse_gain}
\end{figure}

\begin{table}[!t]
\centering
\caption{Neural architectures and training setup used for the nonlinear models. The DAE reconstructs the geometry directly, whereas NLPME reconstructs the original normalized design variables and obtains the geometry through the frozen surrogate of the forward parametric map.}
\label{tab:nn_setup}
\begin{tabular}{lll}
\toprule
Component & DAE & NLPME \\
\midrule
Input & \(\mathbf{d}\in\mathbb{R}^{2352}\) &
\(\mathbf{d}\in\mathbb{R}^{2352}\) \\
Encoder hidden layers & \([1024,1024,512]\) &
\([1024,1024,512]\) \\
Latent dimension & \(N=1,\ldots,32\) &
\(N=1,\ldots,32\) \\
Decoder output & \(\hat{\mathbf{d}}\in\mathbb{R}^{2352}\) &
\(\hat{\mathbf{u}}\in[0,1]^{32}\) \\
Decoder hidden layers & \([512,1024,1024]\) &
\([256,128]\) \\
Surrogate hidden layers & -- &
\([128,512,1024]\) \\
Activation & GELU & GELU \\
Final decoder activation & linear & sigmoid \\
Loss & geometry MSE & geometry MSE \\
Weight decay & \(2.5\times10^{-4}\) & -- \\
Optimizer & Adam & Adam \\
Initial learning rate & \(10^{-3}\) & \(10^{-3}\) \\
Minimum learning rate & \(10^{-6}\) & \(10^{-6}\) \\
Early stopping patience & 500 & 500 \\
\change{Maximum epochs, encoder--decoder} &
\change{\(10^{4}\)} & \change{\(2\times10^{4}\)} \\
\change{Maximum epochs, forward surrogate} &
\change{--} & \change{\(10^{4}\)} \\
\bottomrule
\end{tabular}
\end{table}

The numerical assessment is performed on the valid portion of the bio-inspired glider database described in Section~\ref{sec:test_case}. Each geometry is represented by \(n_g=2352\) geometric degrees of freedom, corresponding to the three Cartesian coordinates of \(784\) aligned surface points.

%Three reduced representations are compared. The first is the linear PME baseline, which provides a geometry-driven reduced space and an analytical backmapping to the original design variables. The second is a deep autoencoder, denoted DAE, used as an unconstrained nonlinear geometric baseline. The third is the proposed NLPME, where the latent variables are decoded into the original normalized design variables and the geometry is reconstructed through the forward parametric map, approximated during training by the frozen surrogate \(\mathcal{S}_{\psi^\ast}\).

The neural setup used for DAE and NLPME is summarized in Table~\ref{tab:nn_setup}. The two models use the same encoder architecture, since the encoder is the component responsible for mapping the geometric representation into the reduced latent variables. In both cases, the input is the standardized shape-modification vector \(\mathbf d\in\mathbb{R}^{2352}\). %, obtained by subtracting the baseline geometry and applying a per-feature \(z\)-score normalization. 
\change{Each geometric component is independently standardized across the sampled configurations using the standard $z$-score transformation.}
The DAE decoder reconstructs the standardized geometry directly, whereas the NLPME decoder reconstructs the original normalized design variables. The latter are constrained to \([0,1]^{32}\) through a sigmoid output activation and are subsequently mapped to geometry through the frozen surrogate of the forward parametric operator.

\change{All neural components are trained using the Adam optimizer with initial learning rate \(10^{-3}\), a reduce-on-plateau scheduler with factor \(0.5\), minimum learning rate \(10^{-6}\), early-stopping patience equal to 500 epochs, and minimum improvement threshold \(10^{-6}\). The DAE is trained for up to \(10^{4}\) epochs for each latent dimension and uses the \(L^2\) weight decay reported in Table~\ref{tab:nn_setup}. In NLPME, the forward surrogate is first trained for up to \(10^{4}\) epochs using a 10\% validation subset for model selection. Once trained, the surrogate is frozen and the NLPME encoder--decoder is trained on the complete valid dataset for up to \(2\times10^{4}\) epochs for each latent dimension. The larger training cap is adopted because the monitored geometric loss continues to decrease beyond \(10^{4}\) epochs. The state associated with the lowest validation loss is retained for the surrogate, whereas the state associated with the lowest training loss is retained for the DAE and the NLPME encoder--decoder.} %No explicit parameter-reconstruction loss, boundary penalty, or latent-space regularization term is included in the present proof-of-concept.

Figure~\ref{fig:nmse_gain} (left) reports the geometric reconstruction error as a function of the latent dimension for PME, DAE, and NLPME, evaluated using the common metric in Eq.~\eqref{eq:nmse}. Both nonlinear methods reduce the reconstruction error much more rapidly than linear PME in the low-dimensional regime. The linear PME baseline requires \(N=8\) reduced variables to reach \(\epsilon\leq5\%\). By contrast, both DAE and NLPME reach the same level with \(N=5\), corresponding to a dimensionality reduction of \(84.4\%\), compared with \(75.0\%\) for PME. Equivalently, at the 5\% reconstruction target, NLPME reduces the number of required latent variables by \(37.5\%\) with respect to PME.

The difference remains significant at the more stringent \(\epsilon\leq1\%\) level. PME reaches this threshold at \(N=15\), corresponding to a dimensionality reduction of \(53.1\%\). The DAE reaches the same threshold at \(N=7\), while NLPME reaches it at \(N=9\), corresponding to dimensionality reductions of \(78.1\%\) and \(71.9\%\), respectively. Thus, NLPME achieves the 1\% reconstruction level with \(40.0\%\) fewer reduced variables than PME. These values are summarized in Table~\ref{tab:thresholds}.

\begin{table}[!t]
\centering
\caption{Latent dimension required by each method to reach the two geometric reconstruction thresholds. Dimensionality reduction is computed with respect to the original 32-dimensional parametric design space.}
\label{tab:thresholds}
\begin{tabular}{lcccc}
\hline
Method &
\(N\) for \(\epsilon \leq 5\%\) &
Reduction &
\(N\) for \(\epsilon \leq 1\%\) &
Reduction \\
\hline
PME   & 8 & \(75.0\%\) & 15 & \(53.1\%\) \\
DAE   & 5 & \(84.4\%\) & 7  & \(78.1\%\) \\
NLPME & 5 & \(84.4\%\) & 9  & \(71.9\%\) \\
\hline
\end{tabular}
\end{table}

\begin{figure}[!b]
    \centering
    \includegraphics[width=0.45\linewidth]{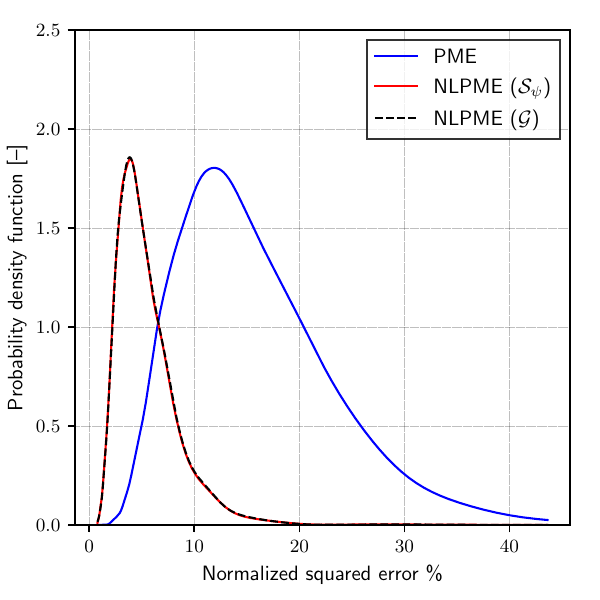}
    \caption{Probability density of the per-sample normalized squared reconstruction error at fixed reduced dimension \(N=5\). PME is compared with NLPME reconstructed through the frozen surrogate \(\mathcal{S}_{\psi^\ast}\) and through the original geometry generator \(\mathcal{G}\).}
    \label{fig:nse_pdf}
\end{figure}

The right panel of Fig.~\ref{fig:nmse_gain} reports the relative reduction of the reconstruction error with respect to PME at equal latent dimension. The nonlinear models provide the largest advantage in the highly compressed regime, where the limitations of the linear subspace are most pronounced. This is the regime of greatest practical interest for design-space reduction, since the objective is not merely to reconstruct the dataset accurately with many variables, but to obtain a compact representation suitable for downstream exploration, surrogate modeling, and optimization.

\change{The DAE reaches the 1\% threshold with fewer latent variables than NLPME, while NLPME retains the parameter-mediated reconstruction pathway. The implications of this comparison are discussed in Section 8.}

The global normalized mean squared error curves in Fig.~\ref{fig:nmse_gain} provide an aggregate measure of reconstruction accuracy, but they do not show how the error is distributed across the sampled design space. For this reason, Fig.~\ref{fig:nse_pdf} reports the probability density of the per-sample normalized squared error defined in Eq.~\eqref{eq:nse_sample} at fixed reduced dimension \(N=5\). This value corresponds to the smallest latent dimension for which NLPME satisfies the target reconstruction level \(\epsilon\leq5\%\). PME is evaluated at the same latent dimension in order to compare the methods at equal compression.

At \(N=5\), PME exhibits a broad distribution of reconstruction errors, with probability density concentrated at larger normalized squared error values. By contrast, the NLPME distributions are shifted markedly toward lower errors. The figure also compares two NLPME reconstruction paths. The first, denoted NLPME\((\mathcal{S}_{\psi^\ast})\), is the surrogate-based reconstruction used during training. The second, denoted NLPME\((\mathcal{G})\), is obtained by feeding the decoded design variables into the original geometry generator. 
%The close agreement between the two distributions indicates that the reported NLPME accuracy is not merely an artifact of the surrogate model; the decoded variables remain consistent with the original parametric geometry-generation pipeline.
\change{The two NLPME distributions are nearly indistinguishable and therefore appear largely superimposed in Fig.~\ref{fig:nse_pdf}. This near-complete overlap indicates that the reported NLPME accuracy is not an artifact of the surrogate model: the decoded variables produce essentially the same reconstruction-error distribution when evaluated through the original parametric geometry generator.}

\begin{figure}[!t]
    \centering
    \includegraphics[width=0.33\linewidth]{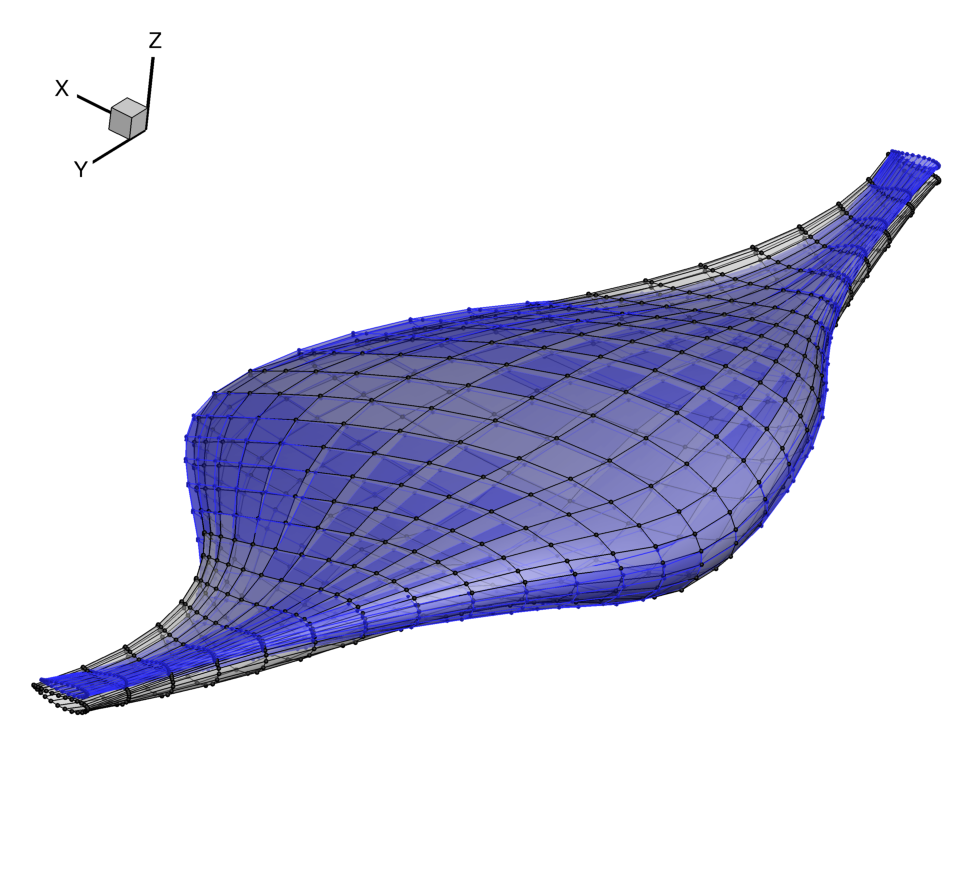}
    \includegraphics[width=0.33\linewidth]{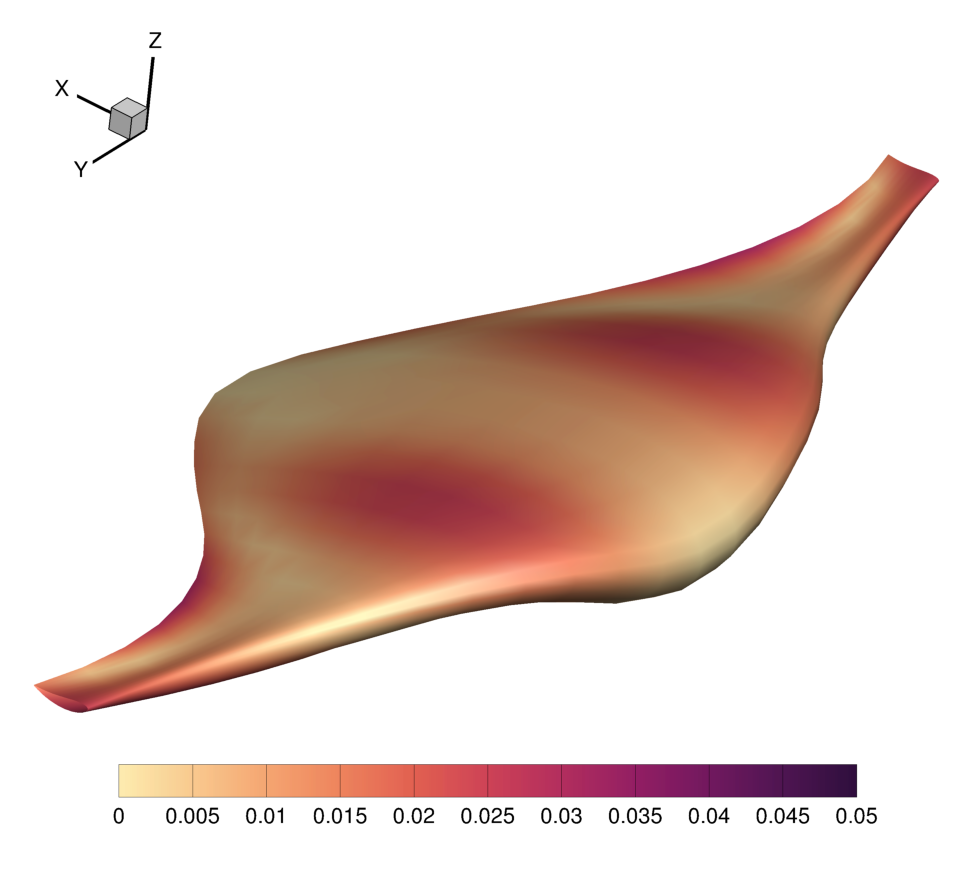}
    \includegraphics[width=0.33\linewidth]{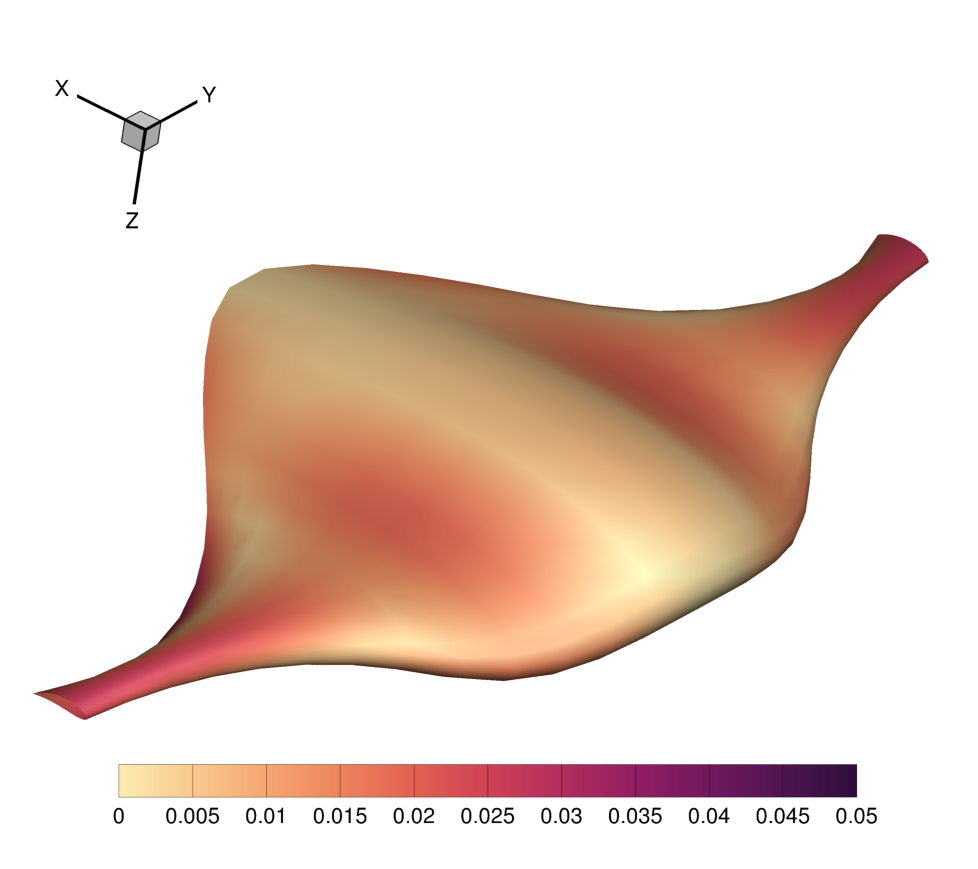}\\
    \includegraphics[width=0.33\linewidth]{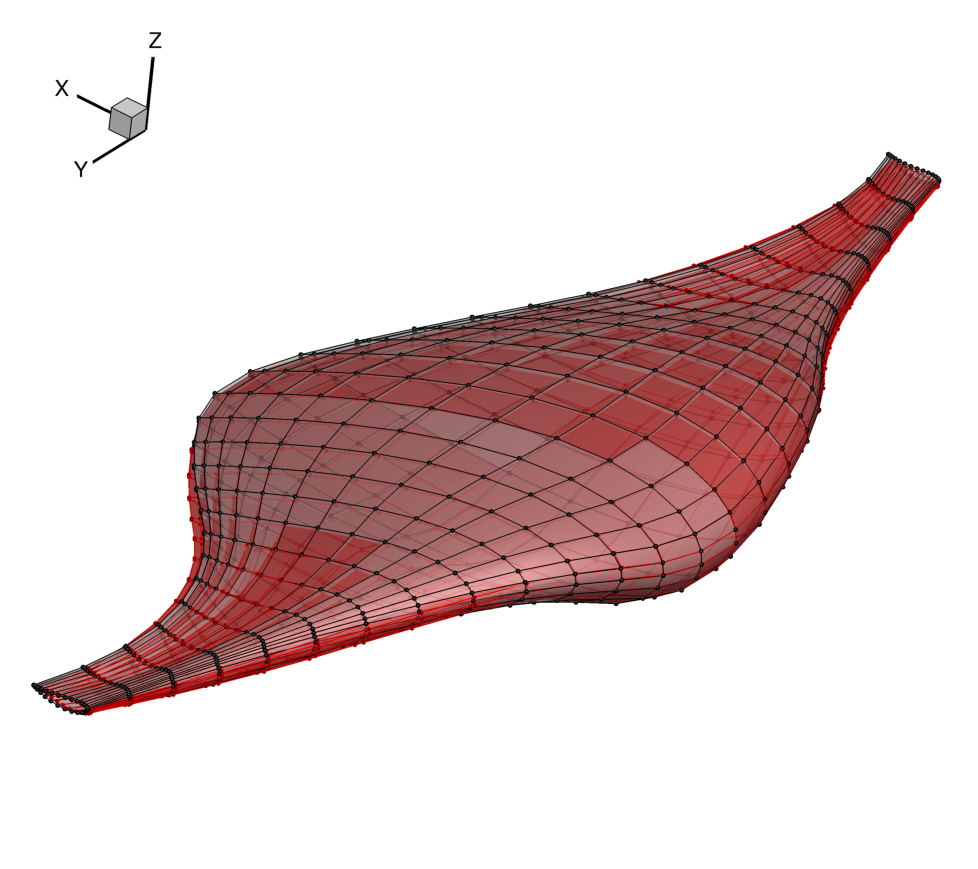}
    \includegraphics[width=0.33\linewidth]{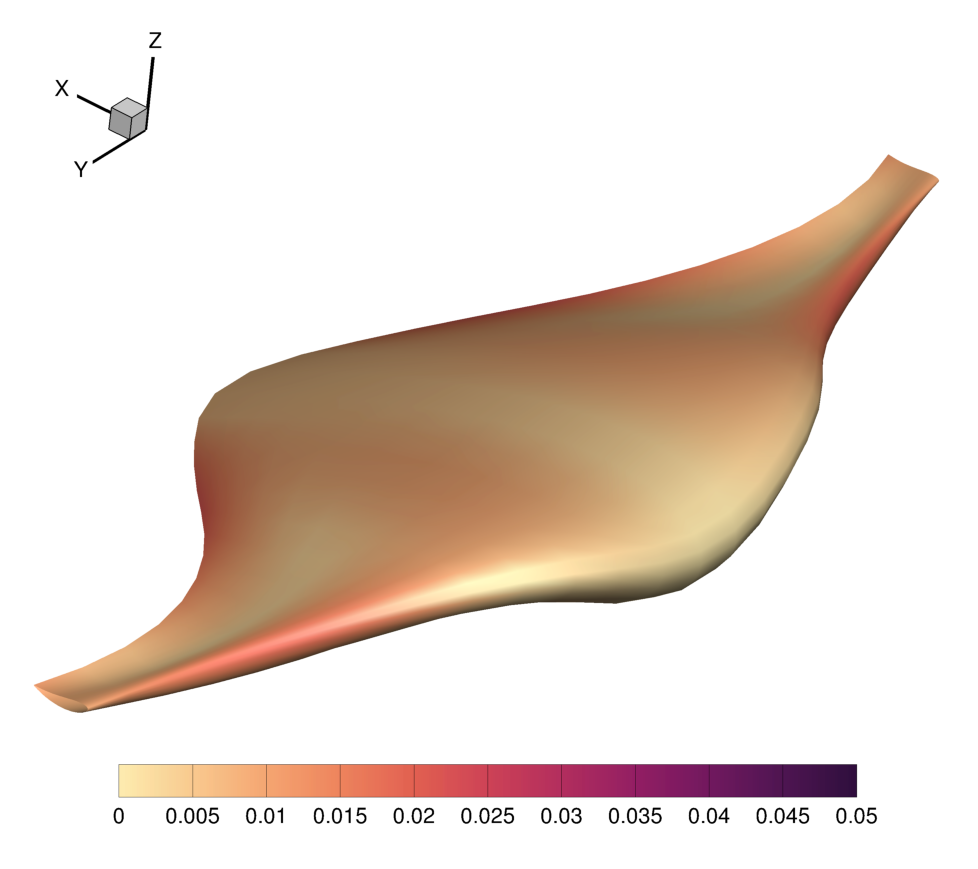}
    \includegraphics[width=0.33\linewidth]{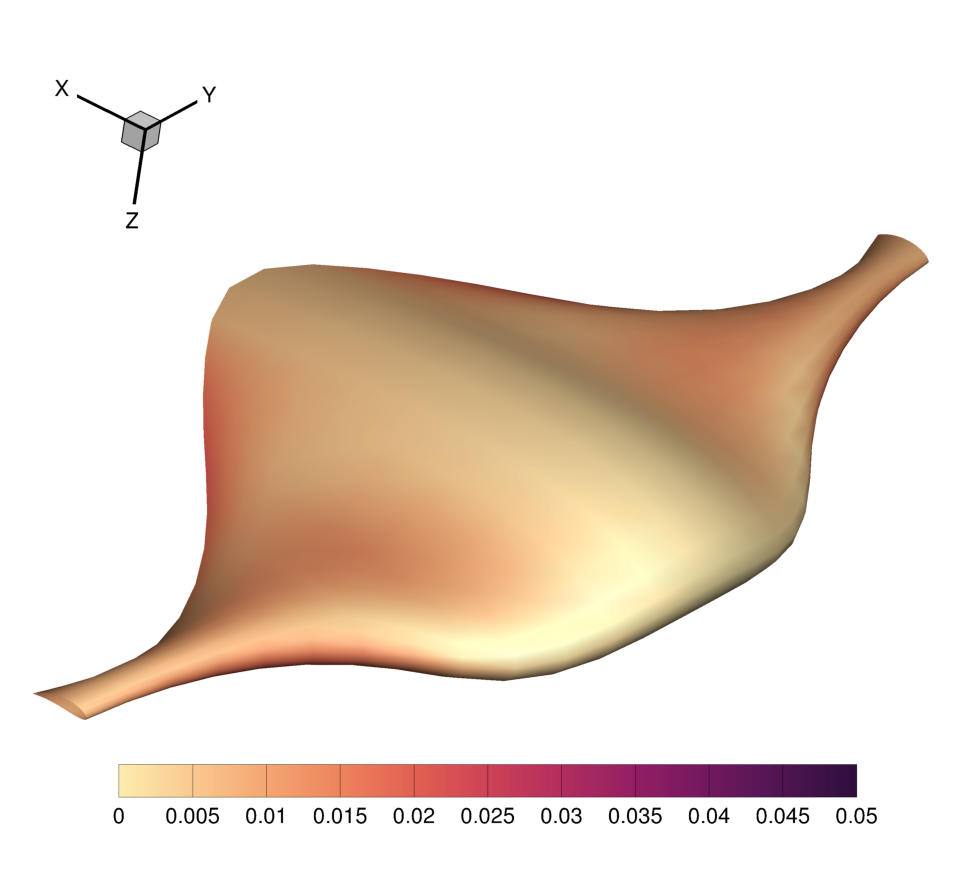}\\
    \caption{Representative reconstruction example at fixed reduced dimension \(N=5\). The selected geometry corresponds to the sample closest to the median of a combined PME--NLPME per-sample reconstruction score, computed as the arithmetic mean of the two normalized squared error values at the same reduced dimension. The surface maps show the local reconstruction error for the two reduced representations (top: PME, bottom: NLPME; left panel: comparison with test geometry in gray; center and right panels: upper and lower surfaces, respectively).}
    \label{fig:reconstruction_example}
\end{figure}

Figure~\ref{fig:reconstruction_example} provides a qualitative assessment of the reconstructed geometries at fixed reduced dimension \(N=5\). The reported geometry corresponds to the sample closest to the median of a combined per-sample reconstruction score, defined as the arithmetic mean of the PME and NLPME normalized squared errors at the same reduced dimension. PME and NLPME are therefore evaluated on the same representative sample and with the same number of reduced variables, so that the visual comparison is performed at equal compression without selecting a case that is favorable to either method. This selection strategy is intended to show a balanced, rather than extreme or method-biased, reconstruction case and provides a visual counterpart to the per-sample error distribution reported in Fig.~\ref{fig:nse_pdf}.

At \(N=5\), the linear PME reconstruction captures the dominant shape variation but exhibits larger local discrepancies. NLPME provides a closer reconstruction of the same geometry while preserving the parameter-mediated reconstruction pathway. The qualitative result therefore reinforces the interpretation of the quantitative metrics: the gain of NLPME is not simply a numerical reduction of the global error, but corresponds to visibly improved reconstruction of a representative geometry from the sampled shape family.

\change{The computational costs of the three representations are summarized in Table~\ref{tab:computational_cost}. The comparison is performed at the smallest latent dimension satisfying the common target \(\epsilon\leq5\%\), so that the methods are assessed at comparable reconstruction accuracy rather than at equal latent dimension. All measurements were obtained on the same workstation, equipped with 12 logical CPU cores and an NVIDIA RTX A5000 GPU, using the complete dataset of 7467 configurations. PME was executed in double precision on the CPU using its NumPy/SciPy implementation, whereas DAE and NLPME were executed in single precision on the GPU using their PyTorch implementations. The reported values should therefore be interpreted as implementation-level wall-clock measurements rather than hardware-independent complexity estimates. Offline cost denotes the shared decomposition for PME, the training of one dimension-specific model for DAE, and the training of the shared forward surrogate plus one dimension-specific encoder--decoder for NLPME. Online cost refers to batch-size-one geometric reconstruction, measured after 100 warm-up evaluations over 10 blocks of 500 repetitions; data transfer, preprocessing, de-standardization, metric evaluation, and file input/output are excluded. For PME, the online path comprises projection onto the latent coordinates and geometric reconstruction; for DAE, it comprises the encoder and geometric decoder; for NLPME, it comprises the encoder, design-variable decoder, and frozen forward surrogate.}

\begin{table}[!b]
\centering
\caption{Observed computational costs at the smallest latent dimension satisfying the geometric reconstruction target \(\epsilon\leq5\%\). Relative costs are evaluated with respect to PME.}
\label{tab:computational_cost}
\begin{tabular}{lccccc}
\toprule
Method &
\(N\) &
Offline cost [s] &
Relative offline cost &
Online cost [ms] &
Relative online cost \\
\midrule
PME   & 8 & 11.96  & 1.00  & 0.02297 & 1.00 \\
DAE   & 5 & 278.10 & 23.26 & 0.19702 & 8.58 \\
NLPME & 5 & 523.10 & 43.75 & 0.27126 & 11.81 \\
\bottomrule
\end{tabular}
\end{table}

\change{For NLPME, the offline value comprises 111.63~s for the forward surrogate, which is trained once and shared among all latent dimensions, and 411.46~s for the \(N=5\) encoder--decoder. The corresponding model-construction costs for the complete sweep \(N=1,\ldots,32\) are 11.96~s for PME, 9020.03~s for DAE, and 12940.76~s for NLPME. PME remains the least expensive method both offline and online, as expected from its single linear decomposition. Therefore, NLPME does not provide a computational speedup over PME or DAE; its contribution is the stronger nonlinear compression obtained while retaining the parameter-mediated reconstruction pathway. Nevertheless, its online reconstruction cost remains below 0.3~ms per configuration for the present test case.}

\section{Discussion}
\label{sec:discussion}

\change{The results should be interpreted primarily in terms of representation structure rather than downstream optimization performance. The numerical evidence indicates that the sampled bio-inspired design space is represented more compactly by a nonlinear latent representation than by linear PME, while preserving the parameter-mediated reconstruction pathway. This gain is consistent with the nonlinear geometric couplings induced by the section-wise parameterization, spatial transformations, and CAD lofting operations of the test case. NLPME therefore extends the representational capability of PME beyond linear subspaces without altering its defining design-oriented logic.}

A related point concerns the structure of the reduced variables. In linear PME, the retained components are naturally hierarchical, since they originate from a truncated generalized PCA basis and can therefore be ranked according to retained variance. In NLPME, by contrast, the latent representation is distributed rather than modal: changing the latent dimension changes the learned representation itself, and the coordinates should not be interpreted as ordered modes with a fixed physical meaning across different bottleneck dimensions. This difference is not a defect of the nonlinear formulation, but a structural consequence of replacing linear modal decomposition with learned nonlinear compression. The practical implication is that NLPME should be evaluated through reconstruction, compression, and backmapping properties, rather than through modal hierarchy alone.

%The use of a differentiable surrogate of the forward parametric map is central to the present implementation, but not to the conceptual definition of NLPME. Methodologically, the surrogate is an implementation-level device that makes the reconstruction chain differentiable. Practically, it avoids repeated calls to an external, potentially expensive, or non-differentiable geometry generator during training. Structurally, it enables gradients of the geometry-consistency loss to propagate through the parameter-mediated reconstruction pathway, thereby making end-to-end optimization of the nonlinear embedding feasible. The comparison between NLPME\((\mathcal{S}_{\psi^\ast})\) and NLPME\((\mathcal{G})\) in Fig.~\ref{fig:nse_pdf} provides a useful consistency check: the surrogate-based reconstruction used for training remains close to the reconstruction obtained by feeding the decoded variables into the original generator. Nevertheless, the surrogate introduces an additional approximation layer, and the present results should therefore be interpreted as assessing the combined effectiveness of nonlinear latent compression and surrogate-mediated geometric reconstruction. Whenever a differentiable and sufficiently efficient exact parameterization is available, the surrogate can in principle be removed and the original generator can be used directly.
The use of a differentiable surrogate of the forward parametric map is central to the present implementation, but not to the conceptual definition of NLPME. Its role is to make the parameter-mediated reconstruction chain differentiable, avoiding repeated calls to an external or non-differentiable geometry generator during training. The comparison between NLPME\((\mathcal{S}_{\psi^\ast})\) and NLPME\((\mathcal{G})\) in Fig.~\ref{fig:nse_pdf} provides a useful consistency check: the surrogate-based reconstruction used for training remains close to the reconstruction obtained by feeding the decoded variables into the original generator. Nevertheless, the surrogate introduces an additional approximation layer, and the present results should therefore be interpreted as assessing the combined effectiveness of nonlinear latent compression and surrogate-mediated geometric reconstruction. Whenever a differentiable and sufficiently efficient exact parameterization is available, the surrogate can in principle be removed and the original generator can be used directly.

The comparison with the DAE baseline should also be interpreted carefully. A DAE reconstructs geometry directly from the latent variables and is therefore free to exploit nonlinear mappings in the geometric space. NLPME solves a more constrained problem, since its decoder must first reconstruct admissible design variables and only then recover geometry through the forward parametric map. The fact that NLPME remains close to DAE in terms of reconstruction accuracy is therefore significant. It shows that much of the nonlinear compression gain over PME can be retained while preserving an explicit link to the original design variables. The relevant comparison is not only the lowest possible geometric reconstruction error, but the combination of compression efficiency and operational backmapping to the parametric model used for geometry generation, meshing, simulation, and optimization.

The present work is intentionally limited in scope. The numerical assessment is carried out on a single parametric shape family. Although the bio-inspired glider case is sufficiently rich for a methodological proof-of-concept, broader validation on additional shape families will be necessary to assess the generality of the proposed extension. The paper also does not address downstream optimization in the reduced nonlinear space. This omission is deliberate: before using NLPME as a reduced-dimensional parameterization for shape optimization, it is necessary to establish that the latent representation is more compact than its linear counterpart and that the associated backmapping remains meaningful and admissible. The current formulation also relies exclusively on a geometry-consistency loss. No explicit parameter-consistency term, latent regularization term, or smoothness-promoting penalty is introduced. This keeps the proof-of-concept focused on the central methodological question, but leaves room for future developments aimed at improving interpolation robustness, regularity of the latent chart, and optimization performance in the reduced space.

Finally, the framework remains restricted to fixed-topology parametric shape families. This is not a limitation of dimensionality reduction per se, but of PME-type methods that require a consistent parametric or pseudo-parametric representation enabling backmapping. Extending the same logic to topology-changing settings, or to cases where the geometric correspondence between samples is not fixed, remains an open research direction.

\section{Conclusions}
\label{sec:conclusions}

% This work introduced a nonlinear extension of parametric model embedding for dimensionality reduction in parametric shape design spaces. The proposed NLPME framework preserves the defining logic of linear PME---geometry-driven latent variables and parameter-mediated reconstruction---while replacing the linear reduced subspace with a nonlinear latent representation. In this way, the method extends the representational power of PME without abandoning its key operational advantage, namely the explicit connection between reduced coordinates and the original design parameterization.

% A central feature of the formulation is that geometry is not reconstructed directly from the latent variables. Instead, the latent representation is decoded into admissible design parameters, and the corresponding geometry is then recovered through a forward parametric map, approximated in the present proof-of-concept by a differentiable surrogate operator. This reconstruction pathway makes the method structurally different from generic nonlinear autoencoders and preserves compatibility with engineering design workflows based on explicit parameterizations.

This work introduced NLPME, a nonlinear extension of parametric model embedding for dimensionality reduction in parametric shape design spaces. The proposed framework preserves the defining logic of linear PME, namely geometry-driven latent variables and parameter-mediated reconstruction, while replacing the linear reduced subspace with a nonlinear latent representation. Geometry is not reconstructed directly from the latent variables; instead, the latent representation is decoded into admissible design parameters and the corresponding geometry is recovered through a forward parametric map, approximated here by a differentiable surrogate.

The numerical assessment on a bio-inspired glider design space shows that the nonlinear extension can achieve a more efficient compression of the sampled shape family than linear PME, especially in the low-dimensional regime where dimensionality reduction is most valuable. The results therefore support the central hypothesis of the paper: when the sampled design space exhibits nonlinear geometric structure, a structured nonlinear extension of PME can provide a more compact reduced representation while maintaining a meaningful parametric backmapping.

The present work is intentionally framed as a methodological proof-of-concept. It establishes the feasibility and potential benefit of nonlinear PME, but does not yet address the use of the resulting latent variables in reduced-space optimization. This next step is a natural continuation of the present study. Once the reduced representation has been shown to be compact, admissible, and operationally meaningful, it can be integrated into simulation-based design optimization workflows in direct analogy with linear PME.

Several extensions follow naturally from the present formulation. A first direction concerns downstream optimization, including the use of NLPME within surrogate-based or Bayesian optimization frameworks. A second direction concerns the training objective itself, for instance through the introduction of additional parameter-consistency, regularity, or interpretability terms. Finally, the broader integration of nonlinear reduced representations with physics-informed or physics-driven embedding strategies remains a promising direction for future research.

%%%%%%%%%%%%%%%%%%%%%%%%%%
\section*{Acknowledgements}
The authors are grateful to the Italian Ministry of University and Research through PRIN 2022 program, project BIODRONES, 20227JNM52 - CUP B53D23005560006. Andrea Serani is supported by the Office of Naval Research grant  N00014-26-1-2164
"Bayesian Exploration and optimization for hull-form Architecture and producibility Modeling (BEAM)" under the administration of Dr. Robert Brizzolara. The work has been conducted within the NATO-AVT-404 Research Task Group on “Enhanced Design Processes of Military Vehicles through Machine Learning Methods”.

%Bibliography
\bibliographystyle{unsrt}  
\bibliography{biblio}

\end{document}